# I am a global citizen. Or am I not? International Business Schools' students and Global Citizenship unified framework & a scoping literature review of the last decade (2013-2022)

**Nikolaos Misirlis[1]**

HAN University of Applied Sciences, the Netherlands[1]

**Abstract**

*I am not Athenian, or a Greek but a citizen of the world (Socrates, 399 BC). Almost two and a half millennia back, Socrates, the Greek philosopher was claiming his global citizenship against his Greek or Athenian one.*

*Today a citizen of the world is considered someone who recognizes and understands the world in its broadest sense as well as his/her place within it. The citizen of the world has an active role in the world and strives for it, recognizing the diversity and striving for a better, more just, and peaceful planet. Today, the need to train young students to become not only worthy scientists but active citizens of the world is even more pressing. More specifically, business school students are now targeting 7 billion potential customers rather than their narrow geographic circle. The need to educate them, therefore, is greater for business students.*

*This review examines the scientific articles of the last decade, approaching the subject through the methodology of the scoping literature review. Starting with the Boolean search "global citizens\*" AND "education" AND ("international business" OR "international business school?") in the ScienceDirect, emerald, and Scopus databases, the review resulted in only scientific journal articles, strictly targeted at tertiary education ONLY of international business schools and ONLY in those articles that study global citizenship. For reasons of up-to-date knowledge, the present literature was content with the final decade.*

*A total of 13 articles are recorded as a result of the aforementioned Boolean search from a total of 216 articles identified in the first phase of the search. The results will help the researchers to acquire the required knowledge base for their research, the academics to incorporate new methods in their teaching and the approach of their students, and the policymakers to adapt the schools' curricula according to the data from the articles present in the literature review.*

## 1. Introduction

Global citizenship refers to the idea that individuals have a sense of belonging to a broader global community and have a responsibility to contribute to the betterment of the world as a whole. It involves recognizing the interdependence of different countries and cultures, and acknowledging that our actions have an impact on others beyond our immediate surroundings [1-3].

Global citizenship in students involves fostering a sense of responsibility towards the world and its inhabitants. It encourages students to think beyond their immediate surroundings and to recognize their role in creating a more just and sustainable future for all [4].

In practice, global citizenship education in schools and universities aims to equip students with the knowledge, skills, and values needed to address global challenges and promote positive change. This includes developing critical thinking skills to analyze complex global issues, as well as promoting empathy and intercultural understanding. In particular, students of international schools around the world feel more of the need to become global citizens, acting like global citizens, working as global citizens, etc [5]. Nowadays the borders of every young worker are not limited to the narrow framework of a country but extend to a global level. But how much are today's students really citizens of the world? How much do they train for it? How ready are academic teachers to train young students to be more ready for the cosmos?

## 2. Methodology

In an effort to understand this framework, this paper is a scoping literature review of academic articles in the last decade. Such reviews are helpful to academics to start building knowledge around a topic [6].The research was limited to this time period in order to record only the most up-to-date results of science. The main question of this research is the following: What are the main results of the scientific community for the readiness of the students and their institutions towards global citizenship?

The result of the research is a complete table with the scientific articles included, and analyzed, the summary for each of them, and a final framework of global citizenship focused on tertiary students of international business schools.

The search for articles was done through Boolean research in the libraries of ScienceDirect and Emerald. The diagram below breaks down the Boolean phrase used ["global citizens*" AND "education" AND ("international business" OR "international business school?")].

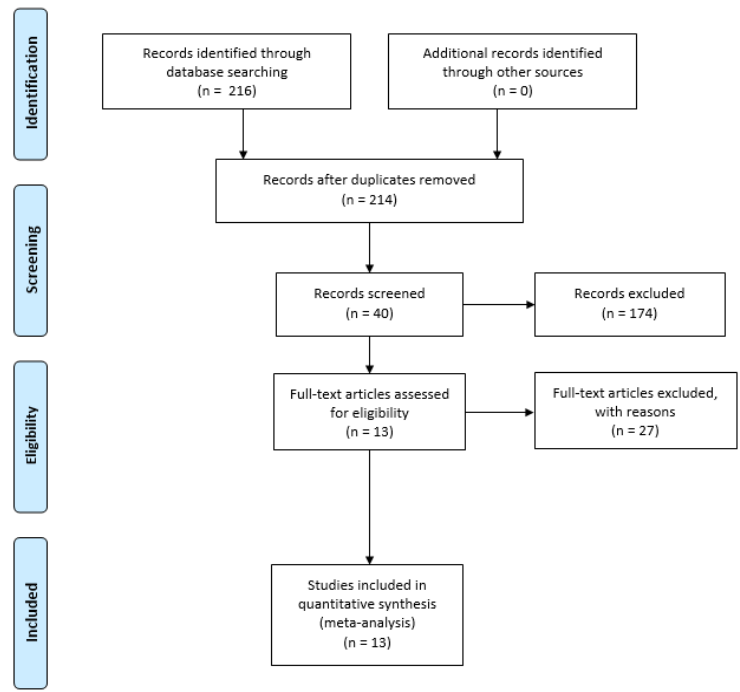

Figure 1: Articles screening

After the first screening, the articles were studied, analyzed and categorized based on the year, and methodology used. The following table summarizes these findings. The most important thing, besides the fact that the table summarizes in one view all the articles into consideration, is the outcome of the main findings of the articles. Keyword analysis and finding analysis has be done in order to create Figure 2. This figure represents a unified framework of global citizenship for students in international business schools. With this figure, researchers, business practitioners and Academics have a clear view of the components they need to incorporate into their research, policies, and academic curricula in order to prepare better the future global citizen/student.

Table 1: Summarize of findings

| Citation | Methodology | Main findings |
|---|---|---|
| Brandauer and Hovmand [7] | Case study | -Use real life case studies and self-assessment are helping students become better global citizens |
| Bulut, Çakmak [8] | Review | -Global citizenship education is suitable to the nature of social sciences.<br>-Global Citizenship education must be evaluated through social networks. In order to accomplish this individuals must be able to reach virtual communication grounds in schools and classes and these grounds must give the |

| Citation | Methodology | Main findings |
|---|---|---|
| | | chance to reach a lot of different cultures and understandings<br>-Technological literacy education must be taken seriously |
| Doerr [9] | Review/ case study | -Students valued immersion yet also learned much in the classroom and by spending time with fellow American and other international students, practices disavowed/discouraged by the discourse of immersion |
| Gulay and Georgiadis [10] | Mixed methods | -foster collaborative research opportunities among several partners, within the "Bridges" project* |
| Roos [11] | Review | -Business schools should be educating future "global citizens", not just "business leaders.<br>-To cultivate responsible actions, the moral and business cases for sustainability are not enough.<br>-A third, governance case based on the old notion of practical wisdom is proposed. |
| Kerkhoff [12] | Qualitative | -the author developed a model related to global readiness students' teaching methods |
| Tyran [13] | - | -Collaborate with people from a variety of disciplines and backgrounds in a developing country through a faculty-led international service learning (ISL) course leads to transform business students to global citizens |
| Dayton, Koster [14] | Review | -Short-term study-abroad programs can offer transformative opportunities for students when intentionally designed as part of curriculum, affecting attitudes toward environmental citizenship and shaping global careers |
| Kirkwood-Tucker, Leierer [15] | Mixed methods | -African-American students possessed a higher level of globalmindedness in the dimensions of efficacy and interconnectedness than their European-American counterparts<br>-higher level of globalmindedness among African-American students in the dimensions of responsibility and cultural pluralism |
| Foster and Carver [16] | Quantitative | -The authors created a toolkit for business school students |
| Woods and Kong [17] | Qualitative | -GCE is an abstraction that has been criticized for reflecting and reproducing (neo)liberal Western values<br>-Inflections of international schools are underpinned by neoliberal and/or cosmopolitan biases |
| Ohajionu [18] | Qualitative research/ interviews | -Need of regular, formal training on internationalization for all academic staff<br>-Lecturers must be more exposed to internationalization |

| Citation | Methodology | Main findings |
|---|---|---|
| Chiang and Chen [19] | Quasi-experimental | - A sustainable design-related course is an effective framework to build the sustainability learning outcomes of international students in a business programme (in the article's case, Chinese students in Thailand) |

A first keyword and findings analysis led to Figure 2, creating a first unified framework, continuously updated with new data, regarding the comprehension of academics and staff towards the proper education of students/ future global citizens.

This framework represents a tool that academics, policymakers, and students should take into consideration and apply to the curriculum in order to follow the latest needs of science and the market. Through these 12 steps, the academia, the businesses, and the staff will be better prepared, adapting their eco-system to global citizenship. Components numbered 01 to 05 refer to the policies that must be taken into consideration, components 6 and 7 refer to the ethics and components 8 to 12 refer to the educational ones.

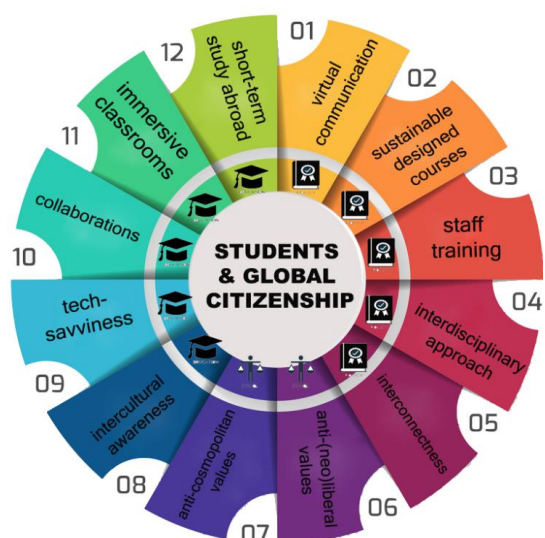

*Figure 2: Global Citizen Framework*

## 3. Conclusions

The importance of global citizenship is today more important than ever. Respectful, diverse, responsible, and collaborative are the students we want in order to build a better future for all. Our students must be individuals who think beyond their personal self-interest, considering the needs of others, both in their own communities and around the world. Universities, by working towards this approach will create eventually a better, more peaceful, and sustainable world for future generations.